# Single-atom doping for quantum device development in diamond and silicon

C. D. Weis[1,2], A. Schuh[1,2], A. Batra[1], A. Persaud[1], I. W. Rangelow[2], J. Bokor[3,4], C. C. Lo[3], S. Cabrini[4], E. Sideras-Haddad[5], G. D. Fuchs[6], R. Hanson[7], D. D. Awschalom[6], and T. Schenkel[1]

[1]Lawrence Berkeley National Laboratory, 1 Cyclotron Rd, Berkeley, CA 94114, USA

[2]Technical University Ilmenau, D-98684 Ilmenau, Germany

[3]Department of Electrical Engineering and Computer Science, University of California, Berkeley, CA 94720, USA

[4]The Molecular Foundry, Lawrence Berkeley National Laboratory, Berkeley, CA 94720, USA

[5]School of Physics, University of the Witwatersrand, Wits 2050, Johannesburg, South Africa

[6]Center for Spintronics and Quantum Computation, University of California, Santa Barbara, California 93106, USA

[7]Kavli Institute of Nanoscience, Delft University of Technology, P.O. Box 5046, 2600 GA Delft, The Netherlands

The ability to inject dopant atoms with high spatial resolution, flexibility in dopant species and high single ion detection fidelity opens opportunities for the study of dopant fluctuation effects and the development of devices in which function is based on the manipulation of quantum states in single atoms, such as proposed quantum computers. We describe a single atom injector, in which the imaging and alignment capabilities of a scanning force microscope (SFM) are integrated with ion beams from a series of ion sources and with sensitive detection of current transients induced by incident ions. Ion beams are collimated by a small hole in the SFM tip and current changes induced by single ion impacts in transistor channels enable reliable detection of single ion hits. We discuss resolution limiting factors in ion placement and processing and paths to single atom (and color center) array formation for systematic testing of quantum computer architectures in silicon and diamond.

---

[1] Email: T_Schenkel@LBL.gov

**1. Introduction**

Continuous miniaturization of classical devices has reached a point where the presence and location of single dopant atoms can strongly affect transport characteristics, even at room temperature [1, 2]. And in the quest for quantum coherent manipulation of information, several implementation proposals are based on the manipulation of quantum states of single atoms in solid state hosts. The latter include spins of donor electrons and nuclei in silicon [3], and spins of nitrogen – vacancy color centers ($NV^-$) in diamond [4, 5]. Techniques for the reliable placement of single atoms into device structures with nm-scale spatial resolution are needed to test the viability of these implementations of quantum computing as well as to understand dopant fluctuation effects in classical device scaling. Bottom-up approaches based on scanning tunneling microscopy and directed self assembly have high (<1 nm) spatial resolution, but are limited to specific surface chemistries, e. g. of phosphorous on clean silicon surfaces [6]. Ion implantation is in general more flexible in the selection of ion species, implant energies and sample materials, but very high resolution ion beams (<20 nm spot size) are primarily available only for gallium ions at a fixed kinetic energy of 30 keV. Moreover, it is highly desirable to be able to image the region of interest without implantation, which precludes the use of ion beams for direct imaging. In our approach to single atom doping, we integrate broad ion beams from a series of ion sources with a scanning force microscope (SFM) [7, 8]. Here, a small (<100 nm) hole in the tip of the SFM cantilever acts as an aperture and defines the beam spot. With this technique we have demonstrated formation of arbitrary patterns in resist layers with feature sizes down to 90 nm [7-10]. Further, we recently demonstrated single atom doping and single atom implantation into transistors with 100% efficiency [11]. We have also addressed a third requirement for single atom device development, namely the retention of dopant arrays and profiles throughout the entire device fabrication process. By studying the diffusion of antimony atoms implanted into silicon with thin oxide layers, we find antimony, a vacancy diffuser, does not show any segregation towards the Si/SiO$_2$ interface [12]. This is in contrast to phosphorus (an interstitial diffuser), which readily segregates to the interface during standard activation anneals [13]. In the following we describe the single atom doping technique, formation of NV-center arrays in diamond and single ion implantation studies of silicon transistors.

**2. Experimental setup and color center pattern formation in diamond**

Figure 1 shows a photograph of the vacuum chamber that hosts the SFM. Broad, low intensity ion beams (~1 pA to 1 µA/mm$^2$) from a series of ion sources enter the implant chamber from the top, where ions with desired mass to charge ratios were pre-selected in a 90º bending magnet. An optical microscope (left) allows pre-alignment of the SFM cantilever (right) to regions of interest on devices and test structures mounted on the sample stage (bottom). The system has a base pressure of ~10$^{-8}$ Torr. The ion sources currently coupled to the SFM are an Electron Beam Ion Trap (EBIT) for high charge state ions (e. g. $^{121}$Sb$^{10-30+}$), and Electron Cyclotron Ion Source (ECR) for medium charge state ions and molecular ions (such as $^{15}$N$_2^+$) and a low plasma density source for singly and doubly charged ions. Test patterns of ion implants can be formed in resist for process development, characterization of tip apertures, or for studies of qubit center formation, by moving the dynamic shadow mask [14] of the pierced tip across the sample during ion bombardment [7-10]. In Figure 2, we show a map of photoluminescence intensities taken at room temperature across a diamond sample that was implanted with nitrogen ($^{15}$N) ions. The map was taken with a long pass filter (>630 nm), which suppresses contributions from light emitted by $NV^0$ centers. Micron scale spots with ensembles of $^{15}NV^-$ centers were formed in ultra pure synthetic diamond (residual $^{14}$N concentration <10 ppb) by implantation of doubly charged, atomic $^{15}N^{2+}$ ($E_{kin}$=14 keV) followed by thermal annealing (800º C, 10 min., in Argon). The background signal from naturally occurring $^{14}$N forming $^{14}NV^-$ centers in random locations is very low in these samples. A key requirement for the integration of $^{15}NV^-$ centers is to achieve center formation with high efficiency [15]. Nitrogen ion doses and Argon ion (28 keV) co-implant doses were varied across the dot pattern to identify optimal local vacancy densities for $^{15}NV^-$ center formation. Figure 3 shows the first results from this study. The PL intensity increases with increasing nitrogen implantation dose. No PL from $^{14}NV^-$ centers was observed for implantation of only argon ions at a dose of $2\times10^{11}$ cm$^{-2}$. Co-implantation of the highest doses of argon ions used here, equivalent to $2\times10^{11}$ to $10^{12}$ cm$^{-2}$, together with implantation of nitrogen ions yields increased PL levels from a given nitrogen implant dose. This demonstrates that the presence of vacancies from the argon co-implant enhances $NV^-$ formation and opens a path to optimization of $NV^-$ formation independent of nitrogen ion implant energy and dose. Detailed studies of this effect of the local vacancy density on $NV^-$ formation probability for a series of implant and annealing conditions are in progress.

**3. Single ion impact detection in transistors**

Single atoms and color centers have to be introduced into single atom devices with high efficiency. The challenge for reliable single ion impact detection can be addressed by detection of secondary electrons emitted by single ions [2, 8], by collection of electron hole-pairs in diodes at low temperature [16], or by detection of current changes induced by single ion hits in the source-drain currents of transistors [11]. Figure 3 shows an *in situ* SFM image of an accumulation channel field effect transistor (aFET) formed in $^{28}$Si. Similar aFETs were recently used for demonstration of spin-dependent neutral impurity scattering, which is a viable mechanism for single spin state detection [17]. A hole was cut into the gate stack of the transistor in a combination of Ga$^+$ ion beam drilling in a Focused Ion Beam (FIB) system and electron beam assisted etching with XeF$_2$. The latter is important to avoid excessive damage to the transistor channel during hole opening. When a pulsed beam of ions impinges onto the device, the source drain current, $I_{sd}$, increases when small ensembles and single ions impinge into the channel region. Figure 4 shows a) the raw $I_{sd}$ data, where the beam of 48 keV Xe$^{6+}$ ions is on during the pulses indicated by the vertical lines. Panel b) shows the smoothed data, and c) the derivative of the smoothed data. Peaks in the derivative signal clearly show (single) ion hits. The short delay between the peaks in c) and the beam pulses is due to signal processing. Transistors are biased in the linear regime with a gate bias of 1.1 V and a source drain bias of 0.1 V and are operated at room temperature. The statistics of hits and no-hits for a series of exposure pulses identifies conditions with single ion hits per pulse. This can be adjusted by tuning beam intensities and selecting appropriate pulse lengths so that at average one ion hit occurs every ~10 pulses (for >99% single ion occupancy per pulse). Beams of noble gas ions are used for tuning, and e. g. antimony ions for formation of (single) dopant atom arrays.

Device currents were found to increase (not decrease) upon ion impacts into channel regions. The data are collected with an inverting pre-amplifier, leading to apparent signal reduction. The mechanism responsible for the single ion induced current changes is well known from radiation hardness studies [11]. Ion hits form electron – hole pairs in the gate oxide. While electrons are quickly swept away, holes remain behind, positive charges enhance the effective gate bias and increases $I_{sd}$. For high doses, however, structural damage becomes important and gradual current reductions are expected.

A key requirement for the development of single-atom devices is the survival of the metallic gates and interconnects of the transistors through the thermal annealing step needed for dopant activation. Use of tungsten for metallization enables this. In Figure 5 we show *I-V* characteristics for a transistor with 2 x 2 µm$^2$ channel area and standard n+pn+ doping configuration (i. e. p-type channel doping and n+ source drain contacts). The significant increase in leakage current due to FIB processing and antimony implantation is restored to the pre-processing performance by the post-implant annealing step (950º C, 20 s, N$_2$, followed by a 10 min. forming gas anneal at 400º C). This enables iterative implantation and electrical characterization (including low temperature transport) of devices with defined numbers of ions in defined dopant patterns for testing of single atom based quantum computer architectures.

Evidently, the method of single ion sensing based on detection of small current changes requires electrical contacting of samples. While this is less universal then e. g. secondary electron emission, the burden it introduces might not be to great. We note that we are able to reliably detect current changes of a few times $10^{-4}$ (with ~1 µA base currents) at room temperature and in relatively large, 2x2 µm$^2$, devices. The disruption of the current through a gate induced two dimensional electron gas by single ion impacts is a very sensitive probe of single ions, and surface currents, or currents through thin sacrificial conductive surface layers are expected to allow sensitive single ion impact sensing. This single ion detection technique can thus be adapted to diamond (and other substrates) by detecting current-upset events in lithographically defined current-channels. Implantation of precise numbers of Nitrogen and Argon ions provides a pathway to understanding the formation efficiency of NV-centers, and the creation of NV-center arrays.

## 3. Discussion and Outlook

The resolution requirements for single atom placement depend on the quantum computer architecture and can vary from a few nm (for coupled NV- centers in diamond [4, 5]) to 100 nm (for Lithium atom qubits in silicon [18]). In our approach, resolution limiting factors are 1) range straggling of ions, 2) the effective beam spot size, and 3) diffusion during activation annealing. Range straggling scales inversely with the ion atomic mass and implant energy, favoring heavy ions like antimony and low implant energies. E. g. straggling in the depth distribution of 25 keV $^{121}$Sb ions in a silicon matrix is <5 nm. Holes with diameters as small as 5 nm have been formed using FIB based drilling and local thin film deposition [19], and recent demonstrations of electron channeling along the hollow cores of multiwall carbon nanotubes suggest that these ultimate beam collimators might enable even smaller effective ion beam spot sizes [20]. Finally, diffusion during required activation anneals is minimal for antimony, and no segregation effects have been found, confirming bulk like diffusivities of ~$6\times10^{-15}$ cm$^2$/s [12], which lead to only minimal dopant movement by a few nanometers even for standard rapid thermal annealing (RTA) conditions (e. g. 1000º C, 10 s). This is in contrast to phosphorus, which segregates during RTA to the SiO$_2$/Si interface [13].

For color center formation in diamond, much more localized wavefunctions of electrons on defect centers lead to coupling lengths scale of only a few nanometers (<10 nm), setting very stringent requirements on beam spot sizes and straggling limits.  These can be reached for pairs of coupled NV$^-$ centers by implantation of $^{15}$N$_2$ molecular ions [21], but conditions for efficient formation of NV$^-$ centers have to be optimized, a task we are addressing through adaptation of defect engineering techniques, analogous to approaches that have been developed for dopants in silicon.

In summary, single ion placement with scanning probe alignment is a universal doping method that enables single atom device development and testing of quantum computer architectures with single atom based qubits.

**Acknowledgments:**   We thank the staff of the Molecular Foundry and the National Center for Electron Microscopy at LBNL for their support.  This work was supported by NSA under contract number MOD 713106A, and by the Director, Office of Science, of the Department of Energy under Contract No. DE-AC02-05CH11231.  Work at UCSB is supported by AFOSR (G.F., D.D.A.)**.**  Support by the National Research Foundation of South Africa is gratefully acknowledged by E. S.-H.

**Figure captions:**

Figure 1: Photograph of the Single-Atom Injector setup. The ion beam enters the vacuum chamber from the top, to the left is an optical viewport, the cantilever is mounted on the right, and the sample stage is on the bottom (with no sample mounted).

Figure 2: Photoluminescence (PL) map of $^{15}NV^-$ centers in pattern with micron scale dots formed by ion implantation with scanning probe alignment using 14 keV $^{15}N^{2+}$ ions and ultra-pure synthetic diamond.

Figure 3: Relative PL intensities from $NV^-$ centers as a function of 14 keV nitrogen ion implantation dose and for a series of argon (28 keV, $Ar^{4+}$) implantation doses ranging from zero to $10^4$ Ar-ions/$\mu m^2$.

Figure 4: *In situ* SFM image of a transistor prepared with a hole in the gate for single ion sensing. The insert shows a schematic of the device cross-section.

Figure 5: Source–drain current as a function of time during pulsed exposure of a transistor to $Xe^{6+}$ ions ($E_{kin}$= 48 keV), a) raw data, b) smoothed data, c) derivative of b).

Figure 6: Room temperature transistor I-V curves before FIB processing, after FIB processing and Sb implantation, and after annealing. The source-drain bias was 1 V.

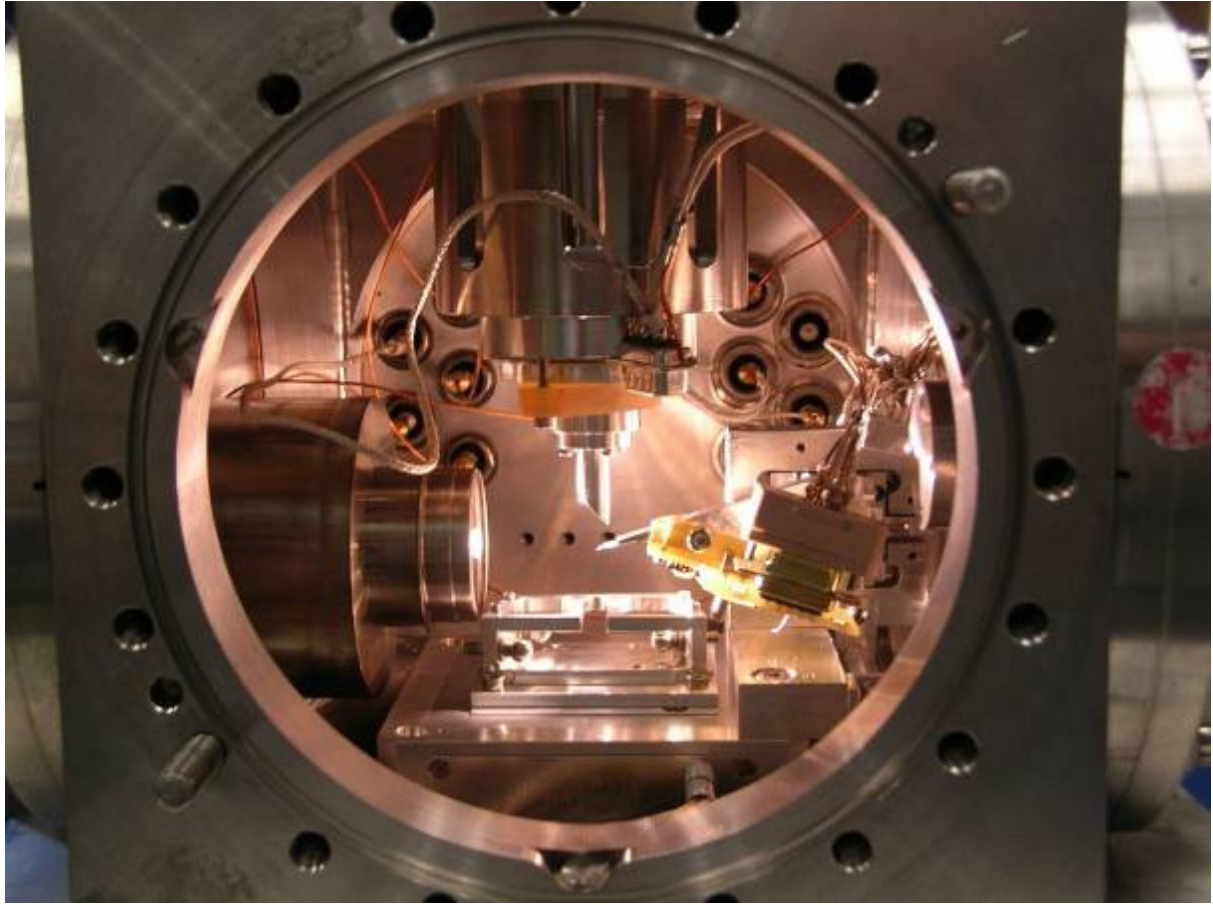

Fig. 1

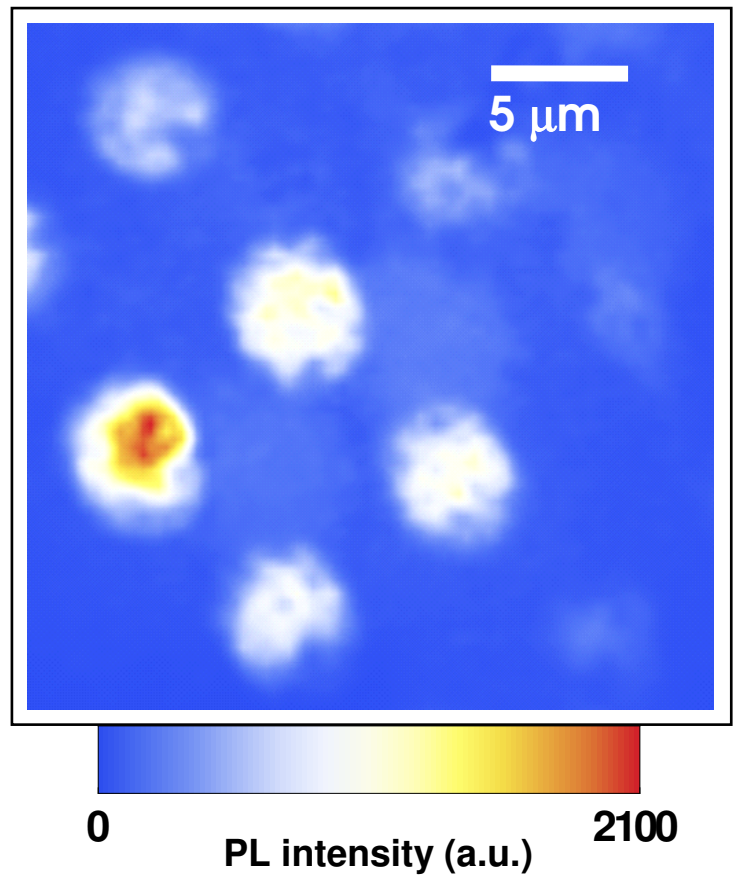

Fig 2

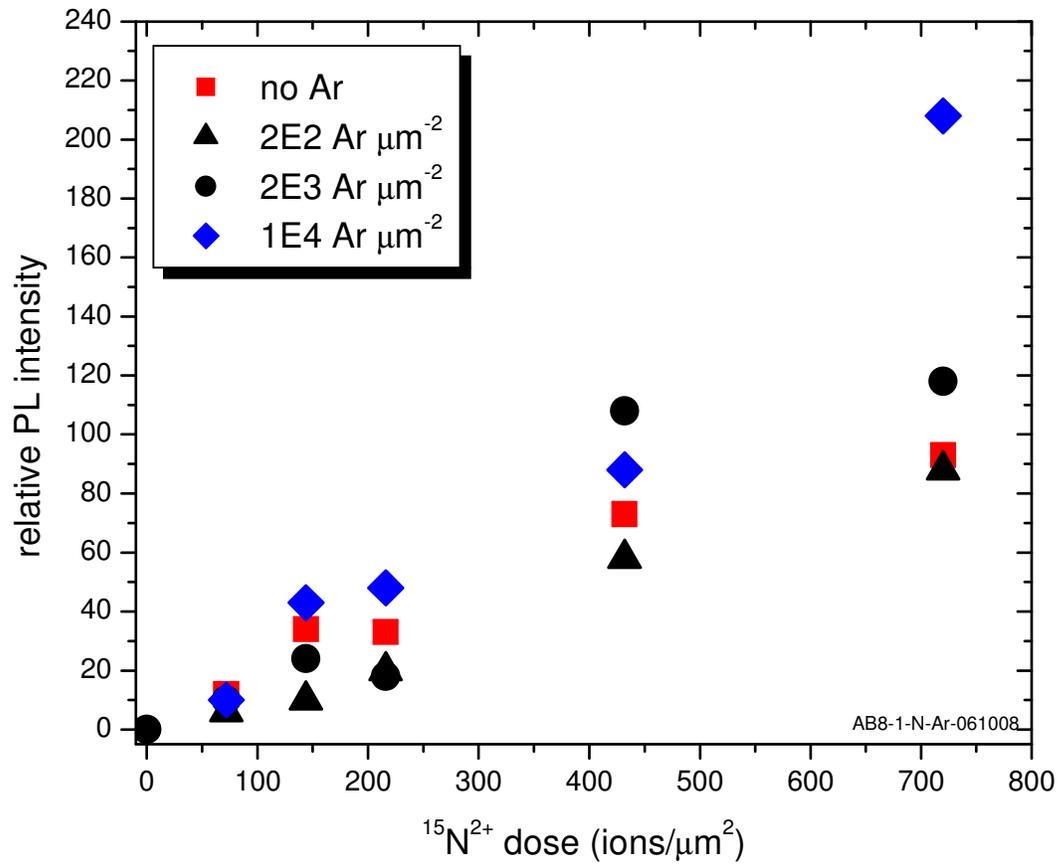

Figure 3

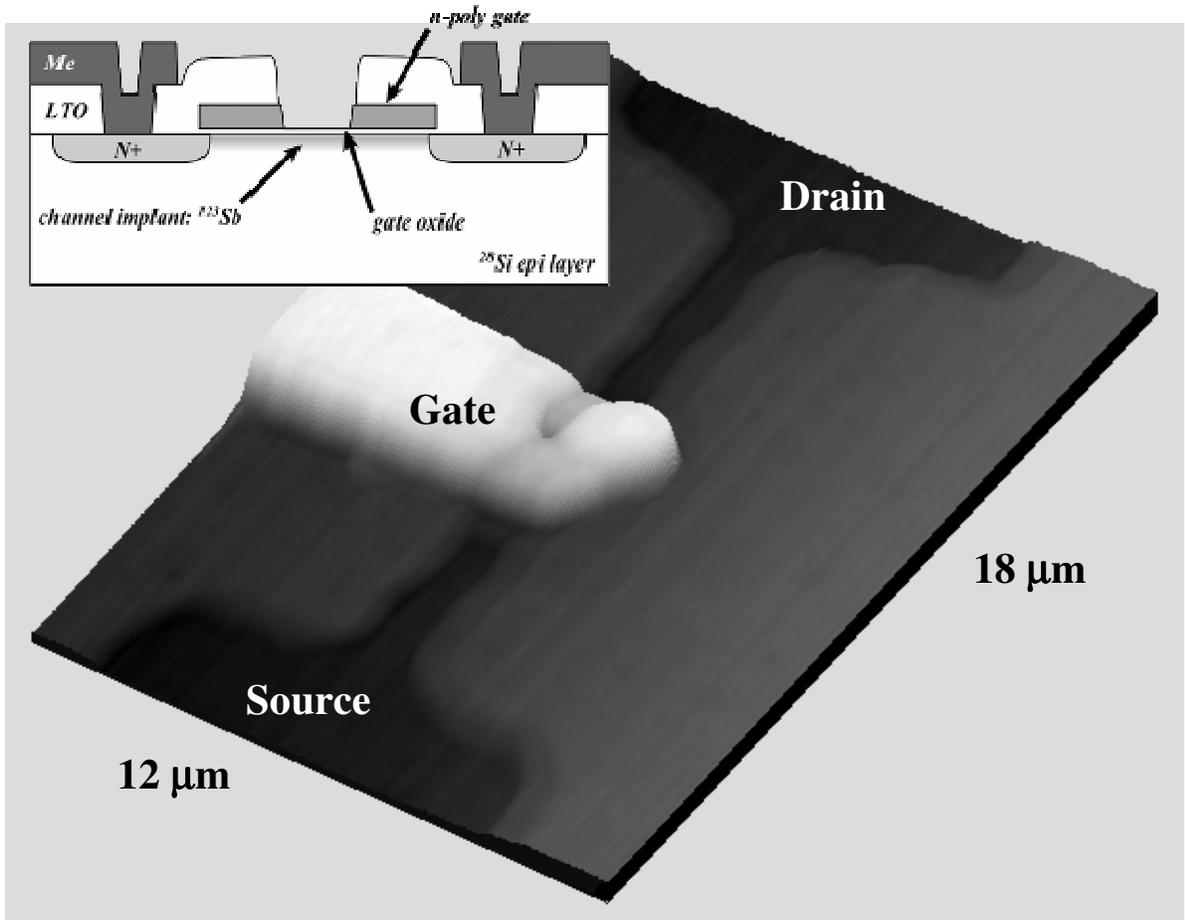

Fig. 4

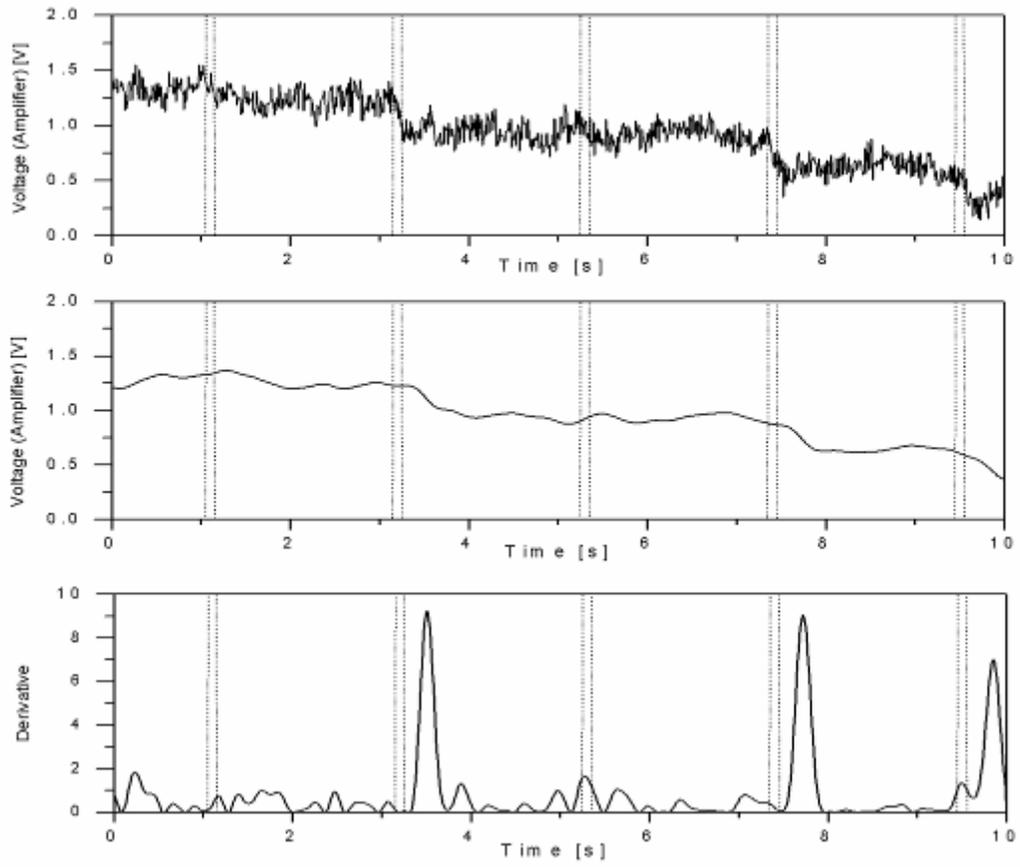

Figure 5 a, b, c

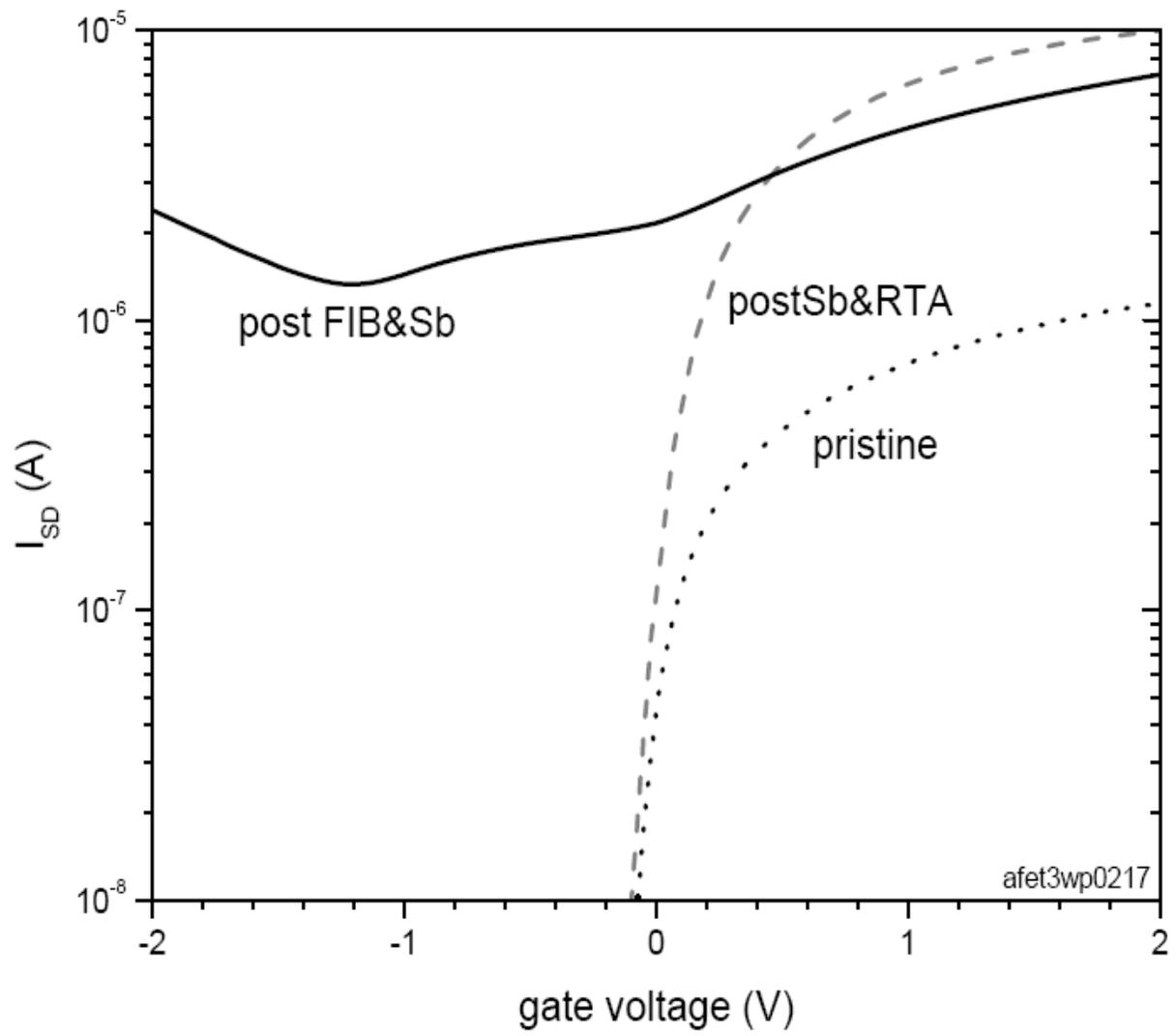

Figure 6